\documentclass[journal]{IEEEtran}

\ifCLASSINFOpdf
\else
\fi
\usepackage{graphicx}

\usepackage{booktabs}
\usepackage[table]{xcolor}
\usepackage[margin=1in]{geometry}
\usepackage{enumitem}
\usepackage{hyperref}

\usepackage{tabularx}
\newcolumntype{C}{>{\centering\arraybackslash}X} % centered version of 'X' col. type

\setlist{nolistsep}
\definecolor{green}{HTML}{66FF66}
\definecolor{myGreen}{HTML}{009900}

\definecolor{MyGreen}{rgb}{0,0.5.2}
\definecolor{MyBlue}{rgb}{0,0.2,1.0}
\definecolor{MyRed}{rgb}{1,0.2,0.2}
\definecolor{MyPurple}{rgb}{0.5,0,1.0}
\definecolor{MyOrange}{rgb}{1.0,0.82,0}
\definecolor{MyBrown}{rgb}{0.65,0.35,0}
\definecolor{MyMagenta}{rgb}{1.,0.0,1.}
\definecolor{MyGrey}{rgb}{0.5,0.5,0.5}

\newcommand{\BB}{BioBlox}
\begin{document}
\title{Bioblox 2.5D --- Developing an Educational Game Based on Protein Docking}
\author{Frederic~Fol~Leymarie,
        William~Latham,
        Guido~Salimbeni,
        Suhail A.~Islam,
        Christopher Reynolds,
        Charlie Cook,
        Luis Armas Suarez,
        Richard Leinfellner,
        and Michael J. E.~Sternberg,% <-this % stops a space
\thanks{FFL, WL, GS, LAS \& RL are with the Department of Computing, Goldsmiths, University of London.}%
\thanks{MJES, SAI, CR \& CC are with the Structural Bioinformatics group, Imperial College London.}%
}%
%%%% The paper headers
%{Shell \MakeLowercase{\textit{et al.}}: Bare Demo of IEEEtran.cls for %IEEE Journals}
% The only time the second header will appear is for the odd numbered pages after the title page when using the twoside option.
% You can use \ifCLASSOPTIONpeerreview for conditional compilation here if you desire.
% If you want to put a publisher's ID mark on the page you can do it like this:
% Remember, if you use this you must call \IEEEpubidadjcol in the second column for its text to clear the IEEEpubid mark.
%%%%% Use for special paper notices
\maketitle
% As a general rule, do not put math, special symbols or citations in the abstract or keywords.
\begin{abstract}
We present the development process of Bioblox2.5D, an educational biology game aimed at teenagers.
The game's content refers to protein docking and aims to improve learning about molecular shape complexity, the roles of charges in molecular docking and the scoring function to calculate binding affinity. We developed the game as part of a collaboration between the Computing Department at Goldsmiths, University of London, and the Structural Bioinformatics group at Imperial College London. The team at Imperial provided the content requirements and validated the technical solution adopted in the game. The team at Goldsmiths designed and implemented the content requirements into a fun and stimulating educational puzzle game that supports teaching and motivates students to engage with biology. We illustrate the game design choices, the compromises and solutions that we applied to accomplish the desired learning outcomes. This paper aims to illustrate useful insights and inspirations in the context of educational game development for biology students.
\end{abstract}

% \fi
\IEEEpeerreviewmaketitle

\section{Introduction}
\IEEEPARstart{E}{ducational} games are designed to help people learn subjects, consolidate concepts and introduce new skills through interactive and empirical approaches. In recent years the use of educational games developed for mobile devices and tablets has gained more traction, exploiting the increased availability of digital tools in classrooms and the latest simulation technologies \cite{barnes2007teaching}. The primary purpose of an educational game is to improve learning, both in and out of the school, so that fun play becomes the means to make that learning process more stimulating, useful and unforgettable \cite{djaouti2011classifying}. According to Squire \cite{squire2011video} educational games should be: (i) a collaborative work of both designers and educators; (ii) entertaining and academically accurate; (iii) both fun and insightful; (iv) based on established game design techniques.

The main goal is to motivate players to learn \cite{prensky2003digital}, while rewards are secondary \cite{michael2005serious}. However, when designing such games, one needs to be aware of some pitfalls which can prevent achieving learning outcomes \cite{prensky2001fun}. For example, the over-simplification of a game can misrepresent the real complexity of the subject; alternatively, the content of the game may not be sufficiently informative.

Another useful practical rule is that an educational game has more chances of success when it is the result of a close collaboration between experts on the subject, who produce and validate the educational contents, and the game development team, who transform the materials through various designs and software programming into an interactive experience of learning. 

In this communication, we present how we followed such ideas to design and develop \textit{\textbf{Bioblox2.5D}}, an educational game in the field of biology. The game is targeted mainly at teenagers between 13 and 17 years of age and to support teaching and motivate students to engage with biology, on the specific topic of \textit{\textbf{protein docking}}.

The game addresses three specific learning outcomes: (i) shape complexity (the size and the representation of molecules in three dimensions), (ii) the roles of charges in the docking mechanism, and (iii) the scoring mechanism of molecular docking. We illustrate the technical choices that we applied to support the learning outcomes. We also describe how we programmed the game so that it matches the requirements provided by the expert team. Version 1.0 of the game is available to download on Android devices from the Google Play store \footnote{\url{https://play.google.com/store/apps/details?id=com.guido.Bioblox2}} and can also be played on-line from the project's website.\footnote{\url{www.bioblox.org}}

%  and funded by the BBSRC \footnote{https://bbsrc.ukri.org/}

\section{Background}
\label{reletedWork}

Several studies have explored the benefit of using computer technology such as used to create digital games to promote meaningful learning and increase students' motivation, as an alternative to traditional teaching. A significant feature of educational games is their positive association with engagement, immersion, favouring concentration all the while generating an enjoyable experience \cite{shernoff2013optimal}.

Engagement and immersion have been commonly believed to be characteristic of an excellent educational game when learning occurs through a process of reasoning upon the simulated world within the game \cite{hamari2016challenging}. Moreover, games can give a break from the normal classroom set-up, which can be particularly appreciated by new generations of students who now frequently make use of digital devices \cite{mayfield2019designing}.

Games such as "The Incredible Machine" \cite{inkpen1995playing} and "Supercharged!" \cite{squire2004electromagnetism} represent successful examples of educational games in Physics. In these games, developers were able to successfully illustrate physics-based contents with simulated dynamics and appealing visualisation, which turned into very engaging games. Not only compelling simulations but also visual and multi-modal environments have been proposed to provide valuable tools for learning \cite{bivall2011haptic}. Interactive computer-based learning materials offer opportunities to help students better understand the concepts by visualising abstract science concepts into a dynamic and concrete experience \cite{srisawasdi2014effect}.

In particular, visualisation and dynamics in educational games can support students in the learning process in biology science because the animated display can help to understand the complexity of biological processes \cite{Baaden2021}. "Cell Craft" is a successful example of an educational and entertaining game to learn about cells, their structure, and how a cell can survive in a hostile environment. This game was effective in increasing students' understanding of science concepts \cite{dunlap2009effects}. "Control of the Cell Cycle" is a different example of a game about the different phases in ordinary cell division, that uses detailed graphics, animations and interactivity \cite{kanyapasit2014development} . 

An important aspect to consider when developing an educational game for biology is that students in that field often have difficulty grasping the complex biological concepts. For example, often students need to memorise numerous terms to better describe the topic of study, which distracts them from a deeper understanding \cite{gutierrez2014development}. Cardona\textit{ et al.} \cite{da2007introducing} have conducted a study on the benefits of an educational game to teach complex content in molecular biology. Olimpo \textit{et al.}  \cite{olimpo2010learning} focus their work on a game that enhances student learning of basic biology terms.  Bhaskar \cite{bhaskar2014playing} found that an educational game could help students improve their understanding of blood grouping to review their existing knowledge. Osier \cite{osier2014board} described how an educational game improves the knowledge of students of genetic terms. 

\subsection{The protein docking problem}

Knowledge of the 3D structure of protein-protein complexes provides major insights into the specificity of interactions, protein function and protein inter-atom networks. Long term benefits of this fundamental knowledge for society include the systematic design of novel pharmaceutical regulators in modern drug design \cite{cheng2012structure,van2017structure} and a molecular-based interpretation of the effect of many mutations in terms of their likelihood to be disease associated.

However there are only structures for about 2,000 complexes between different protein chains because of the difficulties of experimental structural studies. Hence research groups have been developing computational approaches to solve the protein docking problem \cite{Kozakov2017}. The key question for protein docking is whether, starting with the coordinates of two molecules in their unbound \textit{conformations}, one can predict the 3D structure of the docked complex.
Molecular conformation consists in ``any one of the infinite number of possible spatial arrangements of atoms in a molecule that result from rotation of its constituent groups of atoms about single bonds''.\footnote{https://www.britannica.com/science/conformation-molecular-structure} 

Various algorithms have been developed to try to predict the location and rotation of a protein when bound to another protein, a nucleic acid chain or a smaller molecule \cite{tradigo2018algorithms}. It is a complicated task due to a large number of degrees of freedom that need to be calculated combined with the complexity of the proteins structures. Many protein-docking algorithms use a scoring function to predict the binding affinity between two molecules when they are docked together \cite{ogawa2013protein}.

The status of protein docking is assessed in international blind trials, known as CAPRI and CASP \cite{Lensink2017}, where groups predict 3D complexes prior to knowledge of the solution. A major conclusion is that for \emph{ab initio} docking, i.e. when there is no known template to propose solutions, the best algorithms can yield a prediction in about half the submissions that correctly identifies the interface regions together with many of the true atomic interactions. In particular, current algorithms have limited success when there is a substantial conformational change on association.

\subsection{The gamification of science}

Humans have superb spatial reasoning and this has been successfully exploited via crowd sourcing in a range of gamification projects. One of the best known is \textit{Galaxy Zoo} where humans classify the shape of galaxies \cite{GalaxyZoo2008}. In the molecular arena the most well established gamification projects include \textit{FoldIt} \cite{FoldIt10}  and \textit{EteRNA} \cite{Lee2014} which respectively consider folding proteins and the assembly of RNA molecules. Crowd-sourced solutions can outperform computer algorithms or be combined with them in hybrid systems such as ``mixed initiative systems'' \cite{Lai2020Mixed}.

Moreover tracking the approach followed by users has led to the identification of new algorithms. FoldIt, has proved successful in submitted prediction from its players to the related blind trial CASP \cite{Curtis2015}.
A predecessor to the \BB\ project focused on the gamification of protein docking is Udock \cite{Levieux2014}. It starts with classic rigid 3D representations of molecules which can be attached and docked together ``by hand'' via a graphical interface.
We have found that such gamification projects provide useful insights in what may work with a large community of users in terms of interactions when manipulating molecules of various types and representations.
In the next section, we briefly describe the historical development that led us to Bioblox2.5D.
%with the use of appropriate visual graphics, animations and interactivity to facilitate students' scientific conceptual understanding of molecular interactions. 

\section{Bioblox Suite}
\label{suite}

\begin{figure}
 \centering
 \includegraphics[width = 1 \linewidth]{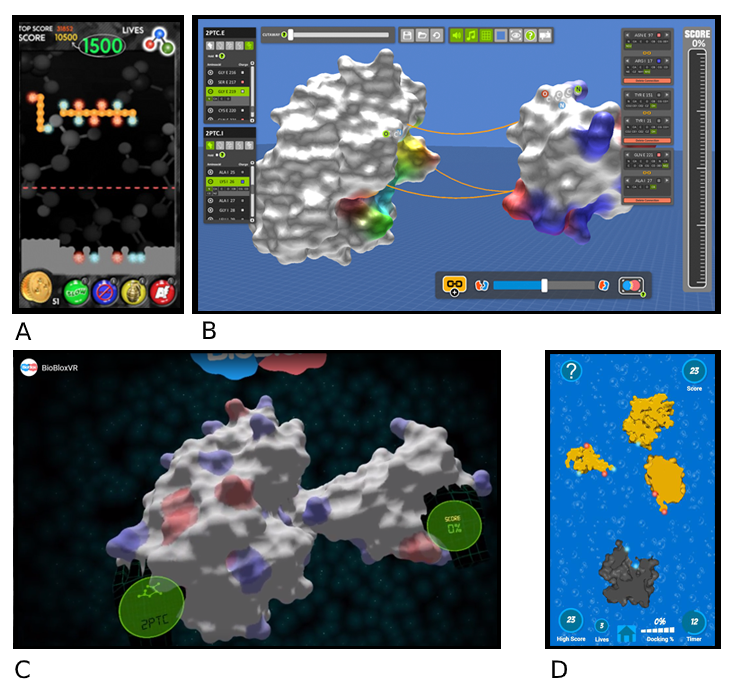}
 \caption{Bioblox Suite: (a) BioBlox2D. (B) BioBlox3D. (C) BioBlox VR. (D) Bioblox2.5D.}
 \label{fig:suite}
\end{figure}

The BioBlox project is the result of a long term collaboration between a computer science group specialising in game computing and a structural biology group specialised in the study of proteins and other biomolecules. BioBlox consists of four freely-available interactive games in which users dock biological molecules, typically proteins (Figure \ref{fig:suite}). 

%begin{description}
%\item[BioBlox3D:]
%\subsection{BioBlox3D}
\textbf{BioBlox3D}, launched in 2016, is a game in which players dock 3-dimensional protein structures. A score is continuously provided based on the quality of the user's alignment. One aim is to provide a platform that can be used by researchers to improve the ability to predict how proteins will interact. Another aim is to evolve it into a citizen science project \cite{follett2015analysis}.  We note that Udock \cite{Levieux2014} offers a similar gaming experience, however with a less elaborate and informative user interface. 

%\item[BioBloxVR:]
\textbf{BioBloxVR} is a version of BioBlox3D, with a simplified interface and ported to Unity to easily run in a virtual reality (VR) platform. It allows users to interact with multiple 3D protein structures, grabbing them with their hands using provided remote controllers. The goal remains to align proteins and maximise a score. Some haptics feedback is provided to the hands of the user via vibrations from the remote controllers. These mimic surface frictions when one molecule is in close contact with another and augment the immersion effect.

%\item[BioBlox2D:]
\textbf{BioBlox2D} has players match both geometry and charges of falling ligands to a receptor at the bottom. The game mechanics are inspired from the classic \textit{Tetris} game  \cite{Demaine2003Tetris}. To progress to higher levels, the players need to answer correctly questions related to molecular science. The game was first launched in 2016 and has been downloaded a few thousand times (from the Apple store and from Google Play). It is designed to run best on a tablet or smart phone.

%\item[BioBlox2.5D:]
  \textbf{BioBlox2.5D} is the latest addition to the suite, and the focus of this communication. It represents an attempt to merge some of the features of BioBlox2D with those of BioBlox3D by using  more realistic shapes. Like its predecessors, it applies simple interactions through touch controls and a user interface that shows the docking scoring.
 %\end{description}

The BioBlox suite is also used in public engagement activities. We use BioBlox2D (since 2016) or BioBlox2.5D (since 2019) to introduce the concept, and people can also play these games on their device while queuing for the 3D or VR experiences. We have participated actively in large science open days and events, including: New Scientist Live in London (2016 and 2017, \url{https://live.newscientist.com}), the Imperial College London Summer Festival (\url{https://www.imperial.ac.uk/festival}) and the Develop conference (the annual main UK based games industry gathering \url{https://www.developconference.com/}), both from 2016 to 2019. We also have presented intermediate results at scientific meetings in the UK and internationally, to help promote the potential of serious games in science as well as in the public at large. The various feedback gained at such events between 2016 and 2019 has helped us in creating and refining BioBlox2.5D.

%\section{Aims of \BB\ 2.5D}
%\label{Aim}
\section{Design of BioBlox2.5D}

Molecular docking represents a significant challenge for educators so that students can comprehend its complexity. It is useful for educators to have a tool that can explain that complexity using graphics, animations and interactivity, which together can overcome the limitation of traditional 2D representations found in a textbook.

To address this issue, BioBlox2.5D aims to be an interactive educational game that tries to improve knowledge of protein interaction by using a combination of 2D and 3D visual representations of the molecules.

%BioBlox2.5D intends to simulate the problem of protein docking with a game design that is a balance between realism and entertainment.
The technical solution is to use 2D slices taken from the 3D model of known proteins as the elements of the game. These slices represent the shape and charges of the proteins.
This solution leverages the fact that 3D images support the learning of such a complex topic, and animations are practical visualisation tools for students that help long-term memory retention \cite{mcclean2005molecular}. 
The game illustrates the docking score through an interactive interface which highlights the docking as a percentage of the perfect dock and alerts the player of weak docking through irritating sound effects. 

\begin{figure}[!htb]
 \centering
 \includegraphics[width = 1.0 \linewidth]{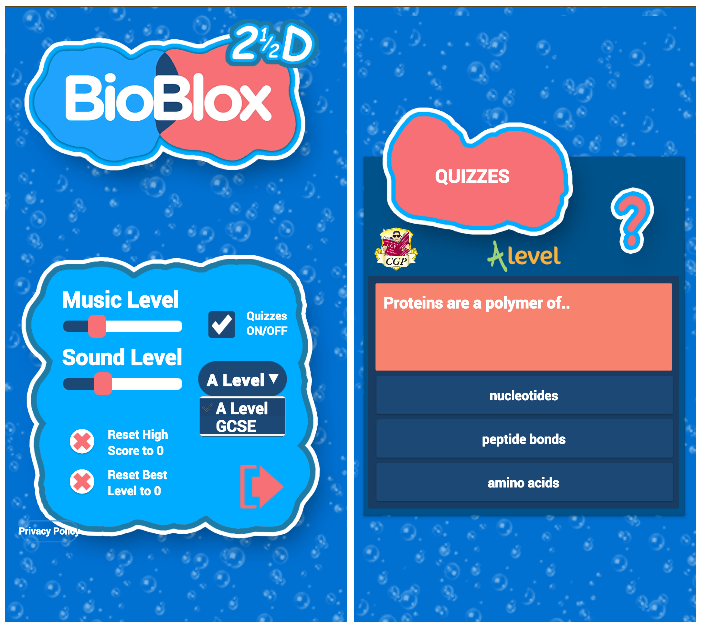}
 \caption{BioBlox 2.5D - Quizzes. Left: Selection of A-Level or GCSE related questions (UK based curriculum). Right: Example of one question.}
 \label{fig:quizzes}
\end{figure}

BioBlox2.5D also achieves engagement through the interaction of the players with simple and intuitive touch controls. The player can also view hints of the proteins and opt to get extra points with quizzes on biology that integrate with the learning aspect of the game (Figure~\ref{fig:quizzes}). At the end of each level, the player can practice their knowledge of biology by answering some questions related to biology and protein docking. The questions are calibrated according to the age of the student and propose three possible answers of which only one is correct. The player receives extra points for a correct answer that adds up to the total score. After the response, either negative or positive, the game shows an explanation of the correct answer.

The quizzes are optional so that the players have the opportunity to pause the game, accessing more challenging concepts or keep the dynamic and fun flow of the game. Having the quizzes as an option follows the idea that well designed educational games have objectives or goals that keep the player in a state of flow and engagement \cite{admiraal2011concept}.
%The player needs to select and drag into place the correct shapes to dock before the time goes over.

\begin{figure}
 \centering
 \includegraphics[width = 1.0 \linewidth]{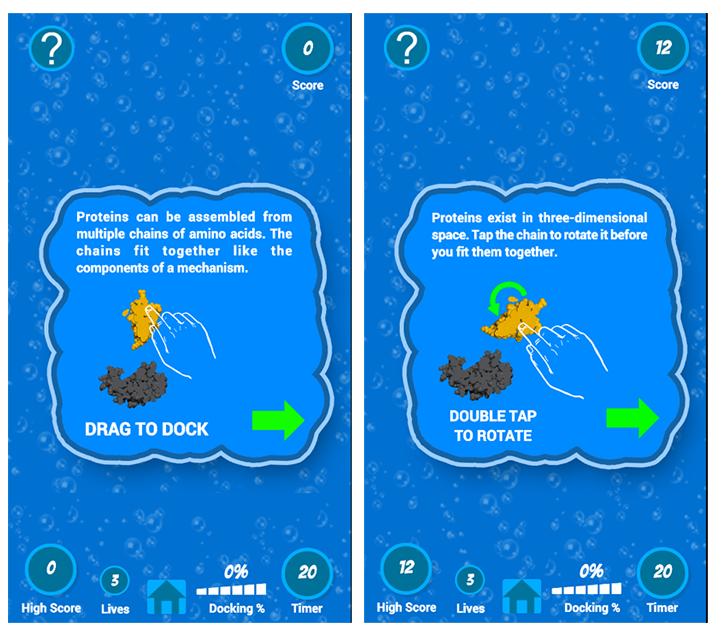}
 \includegraphics[width = 1.0 \linewidth]{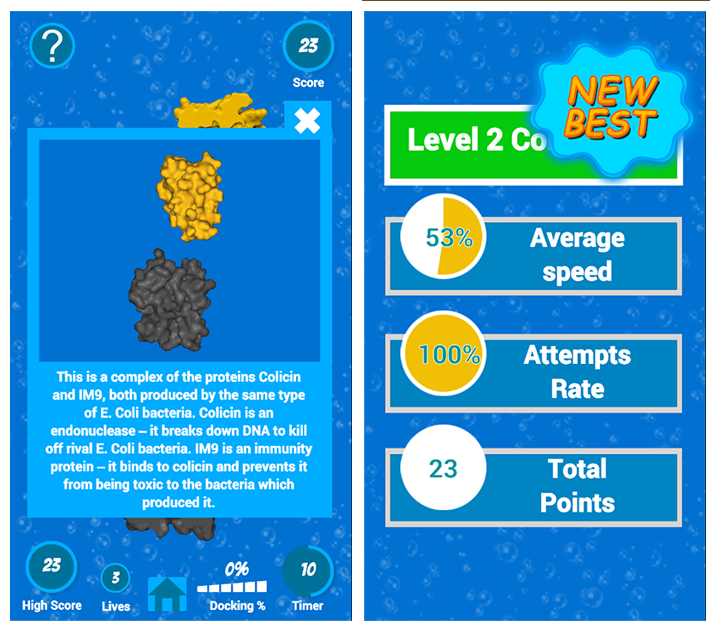}
 \caption{BioBlox 2.5D - Tutorial. Top left: drag tutorial page. Top right: rotate tutorial page. Bottom left: optional page with information about the protein. Bottom right: interface that shows the player's score.
 }
 \label{fig:game}
\end{figure} 

The ultimate goal is to fit the pieces --- the slices of the proteins --- of the puzzle together. To achieve this goal, the player can drag and rotate the slice using touch controls. After pressing the play button  the player starts by reading the first tutorial that explains how to drag the slice of the protein into the docking position (Figure~\ref{fig:game}, top left). Another tutorial shows to the player how to rotate the protein by tapping twice on-screen (Figure~\ref{fig:game}, top right). The game then displays a selection of slices to choose from and points are assigned depending on how quickly the docking is achieved. If the correct match is achieved before the time is up, the player can move to the next round; otherwise, a game life is lost. The player will get additional points if the docking is performed in a short amount of time. A dynamic repulsion force effect indicates an incorrect docking by pushing away the wrong match on the screen. 
There is an optional feature that allows the player to get information about the proteins showed in the game round (Figure~\ref{fig:game}, bottom left). Once all the rounds in the level are completed, an interface shows the total amount of points collected, the average time spent and the precision on selecting the correct protein (Figure~\ref{fig:game}, bottom right). A more detailed description of the levels is in Appendix~\ref{App:levels} 

\section{Learning Outcomes}

Ideally, the game design of an educational game should be implemented as a function of the learning outcomes \cite{baaden2018ten}. We have identified three specific learning outcomes addressed in the game and related to protein docking: (i) complexity of molecular interaction and  representation of proteins shapes and sizes; (ii) importance of charges in the docking mechanism; and (iii) role of scoring function to calculate binding affinity. 
%BioBlox2.5D is a collaborative work of both a team of game developers (at Goldsmiths College) and a team of experts in molecular interactions (at Imperial College London). The result of this collaboration is the design a game that is a balance between entertainment and informative and learning tasks.
This implied that the visualisation and graphics, the interaction design and the scoring were balanced so that the necessary simplifications --- of the shapes, the lack of accurate conformation mechanism, the approximate physics --- do not prevent the delivering of the correct scientific information.  In the following sub-sections, we illustrate how we achieved this balance in various aspect of the game, which are summarised in table \ref{tableSpec}. 

\begin{table}[!htbp]
\centering%
\caption{BioBlox 2.5D - Learning Outcomes \& game design}
\label{tableSpec}
\noindent
%\begin{tabularx}{\linewidth}{XXX}
\begin{tabularx}{\linewidth}{@{} CCC @{}} % to make entries centered
%\begin{tabular}{c|c|c}
%%
\toprule
%\centering%
Learning Outcomes%
& In-game reps%
& In-game mechanics%
\\
\midrule
\midrule
Complexity of molecules interaction and the importance of shapes and sizes in proteins docking.%%
&Molecular mechanism simplified into a puzzle-like game creating the illusion of the 3D structure of protein.%%
&Retain the accurate description of  complex protein structures with an engaging game experience solving dynamic puzzles.\\
\midrule
Importance of charges distribution in the protein docking mechanism.%%
&All salt bridges located between different chains in 3D structures and charges mapped to coloured spheres attached to proteins.%%
&Player needs to match both the shapes and charges to achieve a perfect dock, which is challenging and engaging.\\
\midrule
Role of the scoring function in proteins docking to calculate binding affinity.%%
&Scoring function mapped in the game UI to a numerical value  -- distance to perfect match.%%
&Docking score via an interactive UI highlighting accuracy, with alerts for weak dockings.\\
\bottomrule
\end{tabularx}
%\end{tabular}
\end{table}

\subsection{LO\#1: Molecular complexity, size \& representation.}

Initially, we established the main logic of the game, which we identified as a puzzle-like game. Since proteins interact through a ``lock and key'' mechanism, the puzzle-like game logic is an excellent metaphor to illustrate the challenging task of docking proteins together.

The next step was to design the pieces of the puzzles so that the simplification retains an accurate content definition. Proteins exist in a 3D space and can be assembled from multiple chains of amino acids. The chains fit together like the components of a jigsaw puzzle. We simplified the molecular binding mechanism into a puzzle game using a solution that cuts the 3D shapes of the docking proteins at the known point of contact into 2D slices. This solution allows to retain the academically accurate description of the 3D protein representation and to achieve an entertaining game experience inherent of solving a puzzle. 

\begin{figure*}%[!htbp]
 \centering
 \includegraphics[width = 1.0 \linewidth]{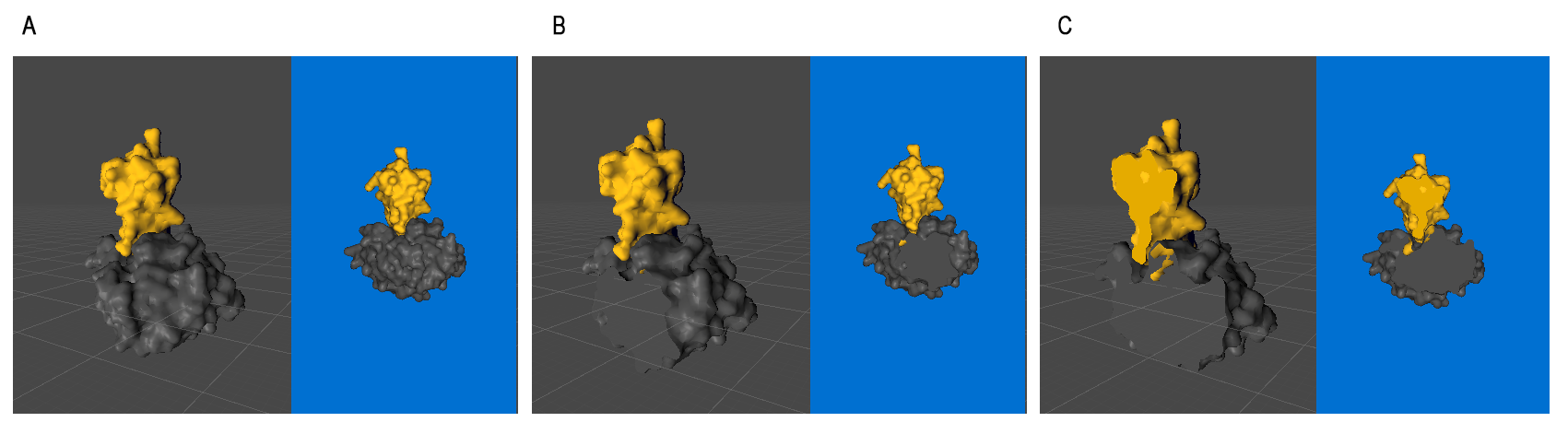}
 \caption{BioBlox 2.5D - Surface representation of two proteins of amino-acids in their docking position (pairs of 3D and 2D views). Left: As downloaded from the RCSB Proteins Data Bank. Middle: Intermediate result of the cutting process. Right: Final representation of the proteins that make explicit the main region of contact.}
 \label{fig:slice}
\end{figure*}

Figure~\ref{fig:slice} briefly illustrates the process that we adopted to create the 2D slices, and it depicts, as an example, the {\it Trypsin} protein complex, which is found in the digestive system. We downloaded the 3D model from the publicly available Protein Data Bank (PDB, entry code 2PTC), which gives access to structural data of biological molecules.\footnote{RCSB - PDB Protein Data Bank -- \href{https://www.rcsb.org/}{www.rcsb.org}.}

The PDB gives data in a specified file format which includes a description and annotation of a protein and its nucleic acid structures, along with atomic coordinates, as well as atomic connectivity. The downloaded 3D model has the chains at their docking position (Figure~\ref{fig:slice}~(A)). Having the chains at their docking position allowed us to digitally cut the 3D model in halves at the exact point of main contact (Figure~\ref{fig:slice}~(B)). Finally, we kept only one-half of each chain that we converted into an image to be used as the piece of the puzzle in the game (Figure~\ref{fig:slice}~(C)).

We repeated the same process for several proteins (listed in Appendix~\ref{listproteins}), producing several cuts to be used as the pieces of the puzzle. This technical solution enables a high-quality graphical representation of the molecules using shadowing and volume rendering that increases the player's immersion in the game and their understanding of a protein shape complexity.

The graphical solution maintains the (illusion of a) 3D visual shape of the proteins, which is a crucial element to the understanding of their biological function \cite{dominguez2003haddock}. Simultaneously, the actual 2D nature of the manipulated image fits well with the puzzle-like game mechanics. 

Our game design illustrates the complexity of molecular interactions by showing realistic renditions of their shapes while letting the user manipulate a simplified --- essentially 2D --- version of the docking mechanism. The player becomes aware of the critical attribute of the proteins that need to be considered to perform the correct dock, such as their relative positions and orientations, compatible size, as well as their complementary charges, which is considered next.

\subsection{LO\#2: Roles of charges in molecular docking.}

Chains of amino acids making proteins are positively or negatively charged. The charges which are located near the outer surfaces of proteins will make them stick together, forming salt bridges to hold the chains in place. The 3D shape and the distribution of the charges are both of great importance to the variety of computational algorithms that try to predict the conformation of binding proteins.
%BioBlox2.5D includes charges in the proteins docking visual representation, being a key element to understand molecular interaction.

\begin{figure}
 \centering
 \includegraphics[width = 1.0 \linewidth]{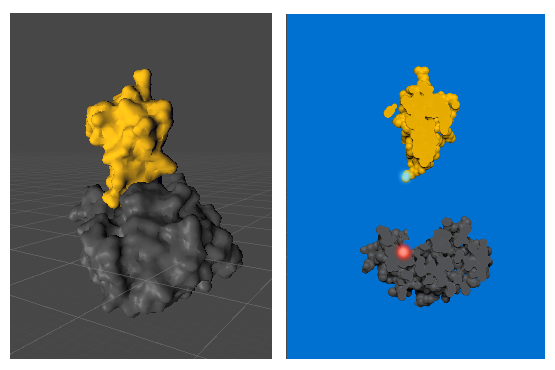}
 \caption{Left: The protein structure first imported inside the 3D modelling software which allows for the separation of chains and annotations of atoms and charges. Right: Once imported in the game the charges are highlighted with coloured spheres.}
 \label{fig:charges}
\end{figure}

\begin{figure}[h]
 \centering
 \includegraphics[width = 1.0 \linewidth]{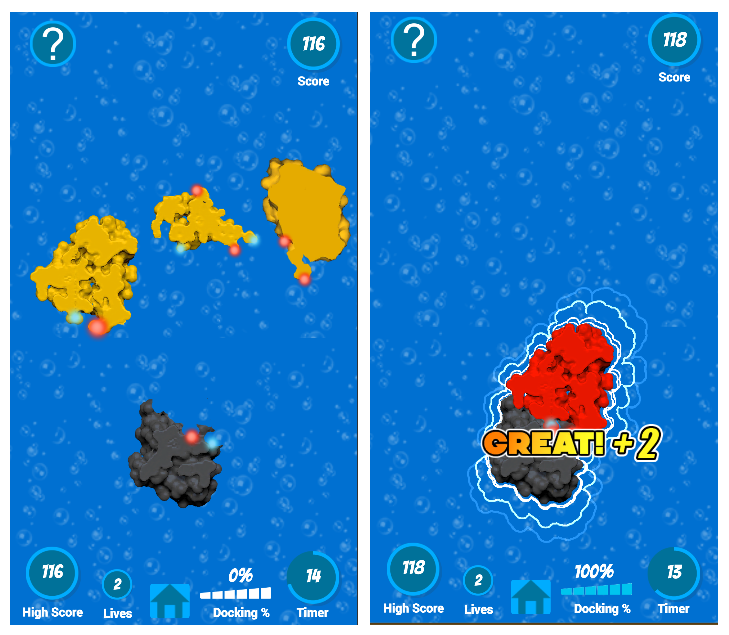}
 \caption{BioBlox 2.5D - The player needs to select the correct shapes that match shapes and charges (A). The winning visual effect of the correct docking (B). Start of the winning 3D animation revealing the 3-dimension of proteins (C). End of the 3D winning animation (D).}
 \label{fig: animation}
\end{figure}

Figures~\ref{fig:charges} and \ref{fig: animation} illustrate the process that we implemented to identify and make the charges explicit along with the shape representation of the proteins in the game. BioBlox2.5D visualises the charges with small coloured spheres placed at the boundaries of the slices, which highlights their positive (blue) or negative (red) electrostatic properties at the correct position.

%To accurately locate the charges, we open a PDB file description of a protein using the Embedded Python Molecular Viewer (ePMV, open source plug-in: \url{http://epmv.scripps.edu/}). We can then visualise the chains that compose the complex protein structure, separate the chains and locate the points of contacts and the charges.

We locate all the salt bridges between different chains in the 3D protein structures using software we
%--- by the team led by one of the co-author at the Department of Bioinformatics of Imperial College London --- 
scripted in Perl, to calculate and sort the contact points by their relative distances, and then annotate the critical points of contact. We then transfer the 3D model of the proteins and the annotations of the charges inside the game. Finally, we used the annotations to place the coloured spheres that represent the charges.

This visual representation of charges fits well with the puzzle-like logic of BioBlox2.5D so that the player needs to match both the shapes and the position of the charges to perform the correct dock. It also represents an opportunity to learn that the availability of 3D proteins structures enables a diligent inspection of their binding site topology and their associated analysis of charge distribution.

\subsection{LO\#3: Scoring function to calculate binding affinity.}

Another important aspect of protein docking is the scoring function that calculates the binding affinity. BioBlox2.5D illustrates this concept using a combination of game interaction and user interface elements that make the learning experience fun and engaging. Proteins realise their functions through binding amongst themselves or to other molecules, and the detailed understanding of the binding procedure is central to biology at the molecular level \cite{du2016insights}. 

%Protein docking simulates the binding of proteins using computational methods. Before the simulated computation, the process involves the selection and preparation of the proteins removing water molecules from the cavity, filling the missing residues and stabilising charges and generating the side chains \cite{ogawa2013protein}. After these steps, the actual docking is computed, and the interactions of the proteins (ligand and receptor) are checked using a scoring function to calculate binding affinity.

Scores depend on the best-selected match. In most docking studies, one crucial problem is the development of an energy function that can rapidly and accurately describe the interaction between the protein and ligand or binding molecule. The success of docking algorithms in predicting protein binding pose is usually measured in terms of a metric comparing the experimentally-observed heavy-atom positions of the molecules and the ones predicted by the algorithm \cite{ding2016assessing}. 

BioBlox2.5D explains the docking scoring function with a numerical value displayed in the user interface of the game and resulting from a simplified accuracy function. The accuracy is calculated as a percentage of the perfect docking position where a 100\% score signifies that all the charges are correctly aligned and overlapping (Figure~\ref{fig: animation}).

%The value of the docking score is set to zero and becomes 100\% accurate when the player achieves the winning dock (Figure~\ref{fig: animation}).
The player attempts to reach 100\% of the docking score in a race against time or one life will be lost. Although in the game each protein is represented as a 2D slice of the original 3D structure, when the correct dock is achieved a 3D animation reveals the actual shape of the protein pairing (Figure~\ref{fig: animation}). This technical solution leverages the fact that 3D images support the learning of such a complex topic, and animations are practical visualisation tools for students that help long-term memory retention \cite{mcclean2005molecular}.

BioBlox2.5D keeps updating and showing the docking score as a percentage of the perfect dock and also alerts the player of a weak docking through repulsion effects.
%(Figure~\ref{fig: ipad} (A - B - C)).
If the dock is only partially accurate, the user interface will show the current percentage of the docking score. %(Figure~\ref{fig: ipad}~(D)).

The information that a player assimilates from an educational game can be very complex, and if everything is presented only in a visual way, the player can lose concentration. \cite{o2010visualizing}. One solution is to simplify the in-game visuals by conveying some of the information through sound \cite{ballweg2016interactive}. For these reasons, we implemented calibrated sound effects that alert the player for the repulsion, the weak score of docking and the winning animation. Even though the user interface of the game displays the current state of the docking accuracy with the percentage visual graphical bar, it is faster and more intuitive for the player to get a sound effect as a response: for a lousy docking score with an irritating sound, and for a good result with a gratifying one.

%\begin{figure*}[!htb]
 %\centering
 %\includegraphics[width = 1.0 \linewidth]{ipad_1.png}
 %\caption{Bioblox 2.5D - touch controls}
 %\label{fig: ipad}
%\end{figure*} 

%\section{Game Development}
\section{Implementation}

Unity3D (\url{www.unity.com}) was the tool chosen to develop this game thanks to its flexibility in managing 2D and 3D animations and the ease it offers to distribute on different platforms. The predisposition of Unity3D for mobile applications permitted an easy implementation of interactions by touch by the player. Although mobile devices, tablets and smartphones are the central distribution platform for BioBlox2.5D, the use of Unity3D has also allowed us to develop a web/PC version of the game (available from the project's website: \url{www.bioblox.org}).

All game dynamics were programmed within Unity3D, while the design of the graphics and animations was done with Adobe's Photoshop and Autodesk's Maya.
%Maya is a software for the animation and modelling of 3D elements while Photoshop is used for the realisation of 2D graphical elements.
We also used the \textit{uPy} plug-in to visualise the 3D molecular structure of the proteins. \textit{uPy} embeds a Python Molecular Viewer and allows a user to load a protein data structure PDB file to analyse and visualise in 3D its content.
%The plug-in was useful to detect the charge positions, the points of contacts, and to separate the chains into two 3D models that we could cut inside Unity3D to produce the pieces of the puzzle.

\section{Conclusion}
We presented the development process of BioBlox2.5D, a digital game that addresses content to be presented in biology classes, focusing on the complexity of molecular docking for students between 13 and 17 years of age. The game is designed around three learning outcomes:  (i) molecular complexity, size and representation; (ii) the roles of charges in molecular docking; and (iii) the scoring function to calculate binding affinity. To illustrate the complexity of proteins docking, we adopted a challenging puzzle-like game logic, and we simplified the 3D forms of proteins into 2D graphics by slicing them at their main point of contact. This solution allows to preserve the correct 3D orientation needed for accurate docking while achieving an entertaining game experience inherent of solving a puzzle. To explain the role of charges in protein docking, we visualised these with small coloured spheres placed at the correct loci on the outline of the slices. These spheres highlight the positive or negative electrostatic properties in the correct position. Finally, to educate about the scoring function that calculates the binding affinity of proteins docking, we added a user interface that displays the game docking score as a percentage resulting from a simplified formula that calculates how well the charges are overlapping.

%We designed the game so that complex biological concepts are presented with visualisations and animations that simplify in a realistic manner the 3D dynamics involved in molecular interaction. To create the game, we developed engaging visual graphics and effects, challenges, connecting goals with the actions of the player. We achieved our goal of developing a game to illustrate the complexity of molecular interactions and proteins docking by showing their realistic shapes even if in a simplified version of the docking mechanism. The player becomes aware of the critical attributes of proteins that need to be considered to perform the correct dock, such as size and conformation, relative positions and orientations. The player can also evaluate the percentage of docking success through an interface that mimics the scoring function used in scientific applications. 

BioBlox2.5D allows students to visualise and manipulate a representation of proteins through time, with text-based definitions accessible at hot-points within the game. There are optional quizzes to test the knowledge on the typical biology questions for students between 13 and 17 years of age (currently based on the UK curriculum). BioBlox2.5D can be used as in-class support to create learning experiences that are attractive, allowing the acquisition and enhancement of knowledge through entertainment and motivation. It can be used as a tool for the teacher to introduce the topic of proteins docking or to augment a typical lesson with an engaging activity. The teacher can make lessons more unforgettable and enhance the usual flow of a lecture or laboratory.

BioBlox2.5D has been successfully used in class since it was launched (mid-2019). The game can either be downloaded for Android devices (freely from the Google play store) or played directly on the web from the project URL (\url{www.bioblox.org}). It has also been used in delivering the Royal Institution Computing Masterclasses to year 9 pupils.\footnote{\url{https://www.rigb.org/education/masterclasses}} We have reports that a majority (circa 80\%) of students keep playing the game during breaks, once it is introduced in class to inform about proteins and molecules (this both for BioBlox2D and BioBlox2.5D).

As future steps, we plan to make more dockings available, add more content for tutorials, tests and information about the science and its uses and potentials. A different version of BioBlox2.5D could focus on protein-ligand docking, where much smaller molecules (ligands) have to fit in protein pockets \cite{Simoes2017}, which is important in drug design. Yet another version would focus on the assembly of multiple proteins, beyond pairs, to construct molecular machines; this would allow for many more game levels to be accessible. Including flexible articulated molecular chains with animations, would allow to demonstrate the more complex conformation mechanisms.

%%% Appendices

% or

\appendices
\section{Levels Design}
\label{App:levels}

The game levels in BioBlox2.5D progressively become more difficult for the game tasks of docking. The first two levels are introduced by two tutorials that graphically explain the dynamics of the game and the touch controls. These first levels have only one protein without charges to be docked. In the next levels, the number of proteins and charges increases. Other factors make the subsequent levels more interesting and stimulating, while learning more about molecular interactions. As the levels progress, the proteins start to appear in movement and in other levels the molecules are pushed by a dynamic force that makes it challenging to find the correct dock. Such dynamics reinforce the peculiar physics and stochastic element that can be observed in nature in molecular interactions.

Table \ref{tableDetails} illustrates in detail all the levels and their respective characteristics. There are currently 7 levels each with several rounds (or sub-levels). For each round the player needs to dock a pair of proteins by choosing against 1 to 3 possible candidates, with the exception of the last level where there are 4 possible candidates. In the initial two levels the charges are hidden and become visible in the following levels. There are three more design features that makes the game more difficult as it progresses. One is the random orientation of the candidates to dock that requires the player to tap on the protein to rotate it before trying to align it into the docking position. Another is a shaking effect that makes one of the proteins move while the player tries to perform a docking. Finally the last two levels add a gravity effect that pushes all docking candidates toward the bottom of the screen, making play harder in selecting proteins and in orienting these to the correct docking position.

\begin{table}[!htbp]
\centering%
\caption{BioBlox 2.5D - Legend: $L$ - level number;  $n_P$ - number of unique proteins to appear in level which can be docked;  $C$ - charge visible (yes or no); $n_R$ - number of rounds per level; $\Theta$ - rotation of protein needed to dock (yes or no); $n_{P/R}$ - number of proteins shown in the current round; $\Delta$ - proteins may be moving on their own; $\cal{G}$ - gravity effect (yes or no). }
\label{tableDetails}
\begin{tabularx}{\linewidth}{XXXXXXXX}
\toprule
$L$ & $n_P$ & $C$ & $n_R$ & $\Theta$ & $n_{P/R}$ & $\Delta$ & ${\cal G}$ \\
\midrule
\midrule
1 & 3 & No & 1 & No & 3 & No & No\\
\midrule
2 & 4 & No & 1 & Yes & 3 & No & No\\
\midrule
3 & 17 & Yes & 2 & No & 3 & No & No\\
\midrule
4 & 10 & Yes & 3 & No & 3 & Yes & No\\
\midrule
5 & 17 & Yes & 3 & Yes & 3 & No & No\\
\midrule
6 & 15 & Yes & 5 & Yes & 3 & Yes & Yes\\
\midrule
7 & 18 & Yes & 5 & Yes & 4 & Yes & Yes\\
\bottomrule
\end{tabularx}
\end{table}

\section{BioBlox2.5D : Proteins currently in use}
\label{listproteins}

All proteins are from the internationally shared resource: the Protein Data Bank (aka PDB). Proteins' 3D coordinates present in the PDB are in open access at the RCSB site (\url{https://www.rcsb.org/}) \cite{Goodsell2020} and the PDBe site (\url{https://www.ebi.ac.uk/pdbe/}) \cite{Armstrong1999PDBe}.
The ten protein pairs currently available are succinctly described below, ordered using their four letter code (from the PDB).

\begin{description}

 \item[1acb: ($\alpha$-chymotrypsin, Eglin):]
    %Docking of the protease alpha-chymotrypsin and eglin.
    Alpha-chymotrypsin is a protease which breaks down other proteins in ingested food and can also cause blood to clot. Eglin is a protease inhibitor produced by leeches to prevent blood clotting by docking to proteases.

\item[1atn: (Actin, DNase I):]
    Actin is a microfilament which provides structure and pathways within cells. DNase I is an endonuclease which breaks down DNA. Actin binds to DNase I to prevent it destroying DNA if the cell is damaged.

 \item[1avx: (Trypsin, Kunitz-type inhibitor):] 
    %Docking of trypsin and a Kunitz-type soybean trypsin inhibitor.
    Trypsin is a protein found in the digestive system which acts as a protease by breaking down other proteins in ingested food. The Kunitz-type soybean trypsin inhibitor is a protease inhibitor found in the coat of soybean seeds that prevents them from being digested when they are eaten by animals.
    
\item[1buh: (CDK2, CksHs1):]
    Docking of the proteins cell division protein kinase 2 (CDK2) and CksHs1. CDK2 is a kinase: it adds a phosphate group to other proteins to provide them with chemical energy. CksHs1 is a kinase inhibitor: it regulates protein activity.
    
\item[1bvn: ($\alpha$-amylase, Tendamistat):]
    %Docking of alpha-amylase and tendamistat.
    Alpha-amylases are proteins found in the digestive system: they break down starch in eaten food into sugars that can be absorbed. Tendamistat is produced by a bacteria that cause plant decay: it prevents the decaying plants being digested if they are eaten.
   
\item[1emv: (Colicin, IM9):]
    %Docking of colicin and immunity protein IM9.
    Both proteins are produced by the same E. Coli bacteria. Colicin is an endonuclease: it breaks down DNA to kill off rival E. Coli bacteria. IM9 is an immunity protein: it binds to colicin and prevents it from being toxic to the bacteria which produced it.
     
\item[1fss: (Fasciculin, Acetylcholinesterase):]
    %Docking of fasciculin and acetylcholinesterase.
    Fasciculin is a toxic protein, aka neurotoxin, found in mamba snake venom. When a snake injects this venom, it interferes with the victim’s nerve transmission and paralyses them. Fascuculin works by docking to and thereby stopping the function of another protein, acetylcholinesterase, which is essential for the transmission of nerve signals.
    
\item[1grn: (Cdc42, Cdc42GAP):]
    %Docking of the cell division control protein 42 (Cdc42) and its activating protein (Cdc42GAP).
    The cell division control protein 42 (aka Cdc42) is involved in regulating cell division. Cdc42 is a GTPase: GTP molecules are the switches which turn on signaling proteins, and GTPases turn the signals off. Cdc42 can bind to its activation protein, Cdc42GAP, which acts as a molecular switch with eglin.
    
\item[2ptc: (Trypsin, inhibitor):]
     %Trypsin is a protein found in the digestive system and acts as a protease by breaking down other proteins in food that have been eaten.
    Trypsin can digest itself in a process known as autolysis. The body has another protein, the pancreatic trypsin inhibitor, which stops autolysis by fitting into the functional centre or active site of trypsin.
    
\item[4kc3: (Interleukin, receptor):]
    %Docking of interleukin and its receptor. 
    Interleukin is a protein involved in the immune system. It belongs to the class of proteins known as cytokines and sends signals between white blood cells so they can coordinate their fight against disease. Interleukin works by binding to another molecule: its receptor.

\end{description}

%%%%%%%%%%End of Appendices 

\section*{Acknowledgements}

This work was supported in part by the BBSRC, Biotechnology and Biological Sciences Research Council, part of UK Research and Innovation, under grant no. BB/R01955X/1.

%%%%%%% Biblio: references section
%\ifCLASSOPTIONcaptionsoff
%  \newpage
%\fi
%\bibliographystyle{IEEEtran}

\bibliographystyle{IEEEtran}
\bibliography{bioblox}

\begin{thebibliography}{10}
\providecommand{\url}[1]{#1}
\csname url@samestyle\endcsname
\providecommand{\newblock}{\relax}
\providecommand{\bibinfo}[2]{#2}
\providecommand{\BIBentrySTDinterwordspacing}{\spaceskip=0pt\relax}
\providecommand{\BIBentryALTinterwordstretchfactor}{4}
\providecommand{\BIBentryALTinterwordspacing}{\spaceskip=\fontdimen2\font plus
\BIBentryALTinterwordstretchfactor\fontdimen3\font minus
  \fontdimen4\font\relax}
\providecommand{\BIBforeignlanguage}[2]{{%
\expandafter\ifx\csname l@#1\endcsname\relax
\typeout{** WARNING: IEEEtran.bst: No hyphenation pattern has been}%
\typeout{** loaded for the language `#1'. Using the pattern for}%
\typeout{** the default language instead.}%
\else
\language=\csname l@#1\endcsname
\fi
#2}}
\providecommand{\BIBdecl}{\relax}
\BIBdecl

\bibitem{barnes2007teaching}
K.~Barnes, R.~C. Marateo, and S.~P. Ferris, ``Teaching and learning with the
  net generation,'' \emph{Innovate: Journal of Online Education}, vol.~3,
  no.~4, 2007.

\bibitem{djaouti2011classifying}
D.~Djaouti, J.~Alvarez, and J.-P. Jessel, ``Classifying serious games: the
  g/p/s model,'' in \emph{Handbook of research on improving learning and
  motivation through educational games: Multidisciplinary approaches}.\hskip
  1em plus 0.5em minus 0.4em\relax IGI Global, 2011, pp. 118--136.

\bibitem{squire2011video}
K.~Squire, ``Video games and learning,'' \emph{Teaching and participatory
  culture in the digital age}, 2011.

\bibitem{prensky2003digital}
M.~Prensky, ``Digital game-based learning,'' \emph{Computers in Entertainment
  (CIE)}, vol.~1, no.~1, pp. 21--21, 2003.

\bibitem{michael2005serious}
D.~R. Michael and S.~L. Chen, \emph{Serious games: Games that educate, train,
  and inform}.\hskip 1em plus 0.5em minus 0.4em\relax Muska \&
  Lipman/Premier-Trade, 2005.

\bibitem{prensky2001fun}
M.~Prensky, ``Fun, play and games: What makes games engaging,'' \emph{Digital
  game-based learning}, vol.~5, no.~1, pp. 5--31, 2001.

\bibitem{shernoff2013optimal}
D.~J. Shernoff, \emph{Optimal learning environments to promote student
  engagement}.\hskip 1em plus 0.5em minus 0.4em\relax Springer, 2013.

\bibitem{hamari2016challenging}
J.~Hamari, D.~J. Shernoff, E.~Rowe, B.~Coller, J.~Asbell-Clarke, and
  T.~Edwards, ``Challenging games help students learn: An empirical study on
  engagement, flow and immersion in game-based learning,'' \emph{Computers in
  human behavior}, vol.~54, pp. 170--179, 2016.

\bibitem{mayfield2019designing}
K.~Mayfield, S.~Cline, A.~Lewis, J.~Brookover, E.~Day, W.~Kelley, and
  S.~Sparks~III, ``Designing a molecular biology serious educational game,'' in
  \emph{Proceedings of the 2019 ACM Southeast Conference}.\hskip 1em plus 0.5em
  minus 0.4em\relax ACM, 2019, pp. 210--213.

\bibitem{inkpen1995playing}
K.~Inkpen, K.~S. Booth, M.~M. Klawe, and R.~Upitis, ``Playing together beats
  playing apart, especially for girls.'' in \emph{CSCL}, vol.~95, 1995, pp.
  177--181.

\bibitem{squire2004electromagnetism}
K.~Squire, M.~Barnett, J.~M. Grant, and T.~Higginbotham, ``Electromagnetism
  supercharged!: learning physics with digital simulation games,'' in
  \emph{Proceedings of the 6th international conference on Learning
  sciences}.\hskip 1em plus 0.5em minus 0.4em\relax International Society of
  the Learning Sciences, 2004, pp. 513--520.

\bibitem{bivall2011haptic}
P.~Bivall, S.~Ainsworth, and L.~A. Tibell, ``Do haptic representations help
  complex molecular learning?'' \emph{Science Education}, vol.~95, no.~4, pp.
  700--719, 2011.

\bibitem{srisawasdi2014effect}
N.~Srisawasdi and P.~Sornkhatha, ``The effect of simulation-based inquiry on
  students’ conceptual learning and its potential applications in mobile
  learning,'' \emph{International Journal of Mobile Learning and Organisation
  1}, vol.~8, no.~1, pp. 28--49, 2014.

\bibitem{Baaden2021}
M.~Baaden, ``{The never-ending quest to understand the shapes and motions of
  molecules},'' \emph{The Biochemist}, 09 2021.

\bibitem{dunlap2009effects}
P.~Dunlap and J.~Pecore, ``The effects of gaming on middle and high school
  biology students’ content knowledge and attitudes toward science,''
  \emph{Studies in Teaching}, pp. 19--36, 2009.

\bibitem{kanyapasit2014development}
P.~Kanyapasit and N.~Srisawasdi, ``Development of digital game-based biology
  learning experience on cell cycle through dslm instructional approach,'' in
  \emph{Proceedings of the 22nd International Conference on Computers in
  Education}, 2014, pp. 857--866.

\bibitem{gutierrez2014development}
A.~F. Gutierrez, ``Development and effectiveness of an educational card game as
  supplementary material in understanding selected topics in biology,''
  \emph{CBE—Life Sciences Education}, vol.~13, no.~1, pp. 76--82, 2014.

\bibitem{da2007introducing}
T.~da~S.~Cardona, C.~N. Spiegel, G.~G. Alves, J.~Ducommun, A.~Henriques-Pons,
  and T.~C. Ara{\'u}jo-Jorge, ``Introducing {DNA} concepts to swiss high school
  students based on a brazilian educational game,'' \emph{Biochemistry and
  molecular biology education}, vol.~35, no.~6, pp. 416--421, 2007.

\bibitem{olimpo2010learning}
J.~T. Olimpo, S.~Davis, S.~Lagman, R.~Parekh, and P.~Shields, ``Learning can be
  all fun and games: constructing and utilizing a biology taboo wiktionary to
  enhance student learning in an introductory biology course,'' \emph{Journal
  of Microbiology \& Biology Education: JMBE}, vol.~11, no.~2, p. 164, 2010.

\bibitem{bhaskar2014playing}
A.~Bhaskar, ``Playing games during a lecture hour: experience with an online
  blood grouping game,'' \emph{Advances in physiology education}, vol.~38,
  no.~3, pp. 277--278, 2014.

\bibitem{osier2014board}
M.~V. Osier, ``A board game for undergraduate genetics vocabulary and concept
  review: the pathway shuffle,'' \emph{Journal of microbiology \& biology
  education}, vol.~15, no.~2, p. 328, 2014.

\bibitem{cheng2012structure}
T.~Cheng, Q.~Li, Z.~Zhou, Y.~Wang, and S.~H. Bryant, ``Structure-based virtual
  screening for drug discovery: a problem-centric review,'' \emph{The AAPS
  journal}, vol.~14, no.~1, pp. 133--141, 2012.

\bibitem{van2017structure}
R.~L. Van~Montfort and P.~Workman, ``Structure-based drug design: aiming for a
  perfect fit,'' \emph{Essays in biochemistry}, vol.~61, no.~5, pp. 431--437,
  2017.

\bibitem{Kozakov2017}
D.~Kozakov \emph{et~al.}, ``The {ClusPro} web server for protein-protein
  docking,'' \emph{Nature Protocols}, vol.~12, no.~2, pp. 255--278, 2017.

\bibitem{tradigo2018algorithms}
G.~Tradigo, F.~Rondinelli, and G.~Pollastri, ``Algorithms for structure
  comparison and analysis: Docking,'' \emph{Encyclopedia of Bioinformatics and
  Computational Biology: ABC of Bioinformatics}, p.~77, 2018.

\bibitem{ogawa2013protein}
T.~Ogawa, \emph{Protein Engineering: Technology and Application}.\hskip 1em
  plus 0.5em minus 0.4em\relax BoD--Books on Demand, 2013.

\bibitem{Lensink2017}
M.~F. Lensink, S.~Velankar, and S.~J. Wodak, ``Modeling protein-protein and
  protein-peptide complexes: Capri 6th edition,'' \emph{Proteins}, vol.~85,
  no.~3, pp. 359--377, 2017.

\bibitem{GalaxyZoo2008}
C.~J. Lintott and other, ``Galaxy {Z}oo: Morphologies derived from visual
  inspection of galaxies from the {Sloan Digital Sky Survey},'' \emph{Monthly
  Notices of the Royal Astronomical Society}, vol. 389, no.~3, pp. 1179--1189,
  2008.

\bibitem{FoldIt10}
S.~Copper \emph{et~al.}, ``Predicting protein structures with a multiplayer
  online game,'' \emph{Nature}, vol. 466, pp. 756--760, Aug. 2010.

\bibitem{Lee2014}
J.~Lee \emph{et~al.}, ``{RNA} design rules from a massive open laboratory,''
  \emph{PNAS}, vol. 111, no.~6, pp. 2122--7, 2014.

\bibitem{Lai2020Mixed}
\BIBentryALTinterwordspacing
G.~Lai, W.~Latham, and F.~F. Leymarie, \emph{Towards Friendly Mixed Initiative
  Procedural Content Generation: {T}hree Pillars of Industry}.\hskip 1em plus
  0.5em minus 0.4em\relax Association for Computing Machinery, 2020. [Online].
  Available: \url{https://doi.org/10.1145/3402942.3402946}
\BIBentrySTDinterwordspacing

\bibitem{Curtis2015}
V.~Curtis, ``Motivation to participate in an online citizen science game: A
  study of {Foldit},'' \emph{Science Communication}, vol.~37, no.~6, pp.
  723--746, 2015.

\bibitem{Levieux2014}
G.~Levieux \emph{et~al.}, ``Udock, the interactive docking entertainment
  system,'' \emph{Faraday Discussions}, vol. 169, pp. 425--441, 2014.

\bibitem{follett2015analysis}
R.~Follett and V.~Strezov, ``An analysis of citizen science based research:
  usage and publication patterns,'' \emph{PloS one}, vol.~10, no.~11, p.
  e0143687, 2015.

\bibitem{Demaine2003Tetris}
E.~D. Demaine, S.~Hohenberger, and D.~Liben-Nowell, ``Tetris is hard, even to
  approximate,'' in \emph{Computing and Combinatorics}, T.~Warnow and B.~Zhu,
  Eds.\hskip 1em plus 0.5em minus 0.4em\relax Springer Berlin Heidelberg, 2003,
  pp. 351--363.

\bibitem{mcclean2005molecular}
P.~McClean, C.~Johnson, R.~Rogers, L.~Daniels, J.~Reber, B.~M. Slator,
  J.~Terpstra, and A.~White, ``Molecular and cellular biology animations:
  development and impact on student learning,'' \emph{Cell Biology Education},
  vol.~4, no.~2, pp. 169--179, 2005.

\bibitem{admiraal2011concept}
W.~Admiraal, J.~Huizenga, S.~Akkerman, and G.~Ten~Dam, ``The concept of flow in
  collaborative game-based learning,'' \emph{Computers in Human Behavior},
  vol.~27, no.~3, pp. 1185--1194, 2011.

\bibitem{baaden2018ten}
M.~Baaden, O.~Delalande, N.~Ferey, S.~Pasquali, J.~Waldisp{\"u}hl, and A.~Taly,
  ``Ten simple rules to create a serious game, illustrated with examples from
  structural biology,'' 2018.

\bibitem{dominguez2003haddock}
C.~Dominguez, R.~Boelens, and A.~M. Bonvin, ``Haddock: a protein- protein
  docking approach based on biochemical or biophysical information,''
  \emph{Journal of the American Chemical Society}, vol. 125, no.~7, pp.
  1731--1737, 2003.

\bibitem{du2016insights}
X.~Du, Y.~Li, Y.-L. Xia, S.-M. Ai, J.~Liang, P.~Sang, X.-L. Ji, and S.-Q. Liu,
  ``Insights into protein--ligand interactions: mechanisms, models, and
  methods,'' \emph{International journal of molecular sciences}, vol.~17,
  no.~2, p. 144, 2016.

\bibitem{ding2016assessing}
Y.~Ding, Y.~Fang, J.~Moreno, J.~Ramanujam, M.~Jarrell, and M.~Brylinski,
  ``Assessing the similarity of ligand binding conformations with the contact
  mode score,'' \emph{Computational biology and chemistry}, vol.~64, pp.
  403--413, 2016.

\bibitem{o2010visualizing}
S.~O'Donoghue \emph{et~al.}, ``Visualizing biological data—now and in the
  future,'' \emph{Nature methods}, vol.~7, no.~3, p.~S2, 2010.

\bibitem{ballweg2016interactive}
H.~Ballweg, A.~K. Bronowska, and P.~Vickers, ``Interactive sonification for
  structural biology and structure-based drug design,'' 2016.

\bibitem{Simoes2017}
T.~Simões \emph{et~al.}, ``Geometric detection algorithms for cavities on
  protein surfaces in molecular graphics: {A} survey,'' \emph{Computer Graphics
  Forum}, vol.~36, no.~8, pp. 643--683, 2017.

\bibitem{Goodsell2020}
\BIBentryALTinterwordspacing
D.~S. Goodsell \emph{et~al.}, ``{RCSB} protein data bank: Enabling biomedical
  research and drug discovery,'' \emph{Protein Science}, vol.~29, no.~1, pp.
  52--65, 2020. [Online]. Available:
  \url{https://onlinelibrary.wiley.com/doi/abs/10.1002/pro.3730}
\BIBentrySTDinterwordspacing

\bibitem{Armstrong1999PDBe}
\BIBentryALTinterwordspacing
D.~R. Armstrong \emph{et~al.}, ``{PDBe: improved findability of macromolecular
  structure data in the PDB},'' \emph{Nucleic Acids Research}, vol.~48, no.~D1,
  pp. D335--D343, 11 2019. [Online]. Available:
  \url{https://doi.org/10.1093/nar/gkz990}
\BIBentrySTDinterwordspacing

\end{thebibliography}


%\end{IEEEbiography}


%\newpage



%\vfill


\end{document}